\documentclass[conference]{IEEEtran}

\usepackage[utf8]{inputenc}
\usepackage[T1]{fontenc}
\usepackage{microtype}
\usepackage{threeparttable}

\usepackage{hyperref}
\usepackage{subfigure}
\usepackage{url}
\usepackage{flushend}
\usepackage{paralist}
\usepackage{siunitx}
\usepackage{booktabs}
\usepackage{tabularx}
\newcolumntype{R}{>{\raggedleft\arraybackslash}X}

\usepackage{dblfloatfix} 

\usepackage{todonotes}
\presetkeys%
    {todonotes}%
    {inline}{}%

\begin{document}


\title{How to specify Non-functional Requirements to support seamless modeling? \\ {\LARGE A Study Design and Preliminary Results} }

\author{\IEEEauthorblockN{Jonas Eckhardt, Daniel M\'{e}ndez Fern\'{a}ndez, Andreas Vogelsang}
\IEEEauthorblockA{Technische Universit\"{a}t M\"{u}nchen\\
Garching b. M\"{u}nchen, Germany\\
\texttt{\{eckharjo,mendezfe,vogelsan\}@in.tum.de}}
}

\maketitle
\begin{abstract}
\emph{Context:} Seamless model-based development provides integrated chains of models, covering all software engineering phases. Non-functional requirements (NFRs), like reusability, further play a vital role in software and systems engineering, but are often neglected in research and practice. It is still unclear how to integrate NFRs in a seamless model-based development.
\emph{Goal:} Our long-term goal is to develop a theory on the specification of NFRs such that they can be integrated in seamless model-based development. 
\emph{Method:} Our overall study design includes a multi-staged procedure to infer an empirically founded theory on specifying NFRs to support seamless modeling. In this short paper, we present the study design and provide a discussion of (i) preliminary results obtained from a sample, and (ii) current issues related to the design. 
\emph{Results:} Our study already shows significant fields of improvement, e.g., the low agreement during the classification. However, the results indicate to interesting points; for example, many of commonly used NFR classes concern system modeling concepts in a way that shows how blurry the borders between functional and NFRs are. 
\emph{Conclusions:} We conclude so far that our overall study design seems suitable to obtain the envisioned theory in the long run, but we could also show current issues that are worth discussing within the empirical software engineering community. The main goal of this contribution is not to present and discuss current results only, but to foster discussions on the issues related to the integration of NFRs in seamless modeling in general and, in particular, discussions on open methodological issues.

\end{abstract}

\IEEEpeerreviewmaketitle

\section{Introduction}

The increasing complexity in software and system development projects results in a demand for expressiveness, modularity, reusability, and analyzability of modeling and specification approaches. Model-based development is the key to meet this demand as it allows to abstract from implementation details and to increase the overall level of abstraction. Yet, model-based development alone does not solve anything as various models still have to be integrated into a holistic chain. In response to this, the idea of seamless model-based development emerged~\cite{broy2010seamless}. Seamless modeling aims at elaborating integrated chains of models covering all phases from requirements engineering to system design and verification. 
The central ingredient of seamless model-based development is a system model that provides the theoretical framework interconnecting all models.

 Non-functional requirements (NFRs) further play a vital role in software and systems engineering. There is much work available in the field of NFRs characterizing single classes of NFRs and classifications as well, such as security and reusability. Yet, we are still far from having a common understanding on the notion of NFRs, let alone from having commonly accepted and integrated taxonomies for NFRs~\cite{iso9126,glinz2007non,chung2009non} that go beyond a rather abstract level, or even an integration of NFRs in a common system model. In fact, the integration of NFRs and model-based development forms a high priority scope of current research projects that aim at better understanding how practitioners integrate NFRs in context of model-based development and, in particular, what problems they experience~\cite{NFR2015,svensson2013investigation}.\looseness=-1
 
Therefore, it currently remains unclear how to integrate NFRs in seamless model-based development, as the integration in a common system model is not in scope of  available contributions. This forms the objective of our ongoing research. In the long run, we want to provide an approach for specifying NFRs such that they can be integrated in a common system model, and thus, supporting seamless model-based development that also takes into account the specification of NFRs.

 To reach this goal, we designed an overall study that starts with analyzing how practitioners specify NFRs. This literature-agnostic view allows for getting an overview of the information and structure necessary to specify NFRs sufficiently suitable for subsequent development activities (even if not integrated). Another reason why we base our work on practical data is that we want our resulting theory to emphasize the practical impact rather than the theoretical one alone. Having understood the basic constructs used to specify NFRs in practice, we analyze in a second step the relation between classes of NFRs and the various system modeling dimensions. We use this classification to elaborate, in a third step, a theory on specifying NFRs in context of seamless development, before eventually disseminating and evaluating our theory again in practical contexts. \looseness=-1
 
  In this paper, we present our overall design and discuss current results and methodological issues arising from the preliminary analysis of a sample. Please note that the primary aim of this paper is not to present the results alone (as the study design still might be subject to change), but to foster discussions and exchange ideas on this difficult area characterized by various (empirical) challenges.

\section{Overall Study Design}
The goal of the overall study is to analyze natural language NFRs taken from industrial requirements specifications in order to understand how classes of NFRs relate to existing system modeling dimensions. This serves as a basis for developing a theory for the specification of NFRs to support seamless modeling. 

\subsection{Research Questions}
To reach our goal, we formulate the following research questions (RQs):

\textbf{RQ1:}
{\itshape What classes of NFRs are documented in practice and what is their scope?} This RQ examines the state of practice, i.e., what classes of NFRs are actually documented and whether they refer to the context, to the system or to a sub-system.

\textbf{RQ2:} {\itshape How do classes of NFRs relate to existing system modeling dimensions?}
In this RQ, we lay the foundation of the later theory building: we analyze how (and if) classes of NFRs relate to specific system modeling dimensions.

{\color{gray}
\textbf{RQ3:} {\itshape How can NFRs be specified to support seamless modeling?}
For those NFRs that are related to system modeling dimensions, we build in a third step a theory on how to specify NFRs to support seamless modeling. 

\textbf{RQ4:} {\itshape What are the limitations of specifying NFRs in a seamless context?} In a last step, we analyze the limitations of the resulting theory which we also plan to use for further adjustments of the concepts captured in the theory.
}

\begin{figure}
\centering
\includegraphics[width=0.75\columnwidth]{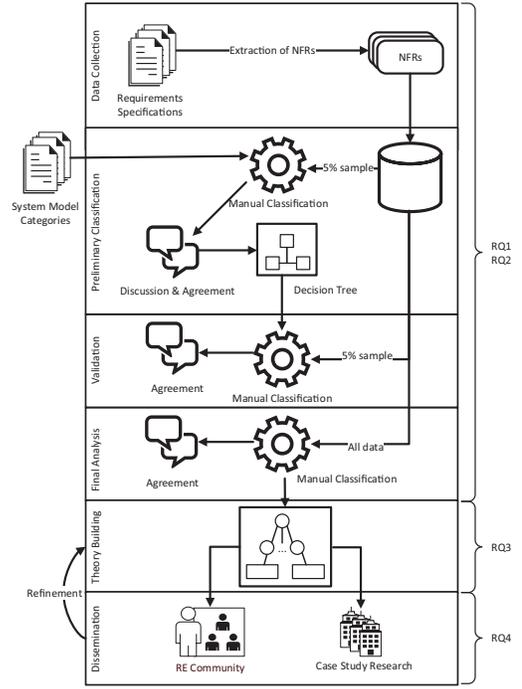}
\caption{Overview of the study design and the relation of the RQs to the steps.}
\label{fig:process}
\vspace{-1em}
\end{figure}

Figure~\ref{fig:process} depicts an overview of the overall study design. First, we perform a preliminary classification of a sample (approx. 5\%) to see how the individual classifications align, followed by a discussion with an agreement. Based on this discussion, we build a decision tree for the classification to make the classification more transparent. Using this decision tree, we validate the classification on another random sample (approx. 5\%), followed again by a discussion with an agreement. Finally, to answer RQ1 and RQ2, we analyze the whole data set (not in scope of the paper at hands). While RQ1 and RQ2 are concerned with the collection and classification of NFRs from concrete requirements specifications, RQ3 is concerned with theory building before we evaluate the resulting theory again in a practical (industrial and academic) context (RQ4). 

Please note that in this paper, we provide insights into the current data analysis that concerns RQ1 and RQ2.\looseness=-1

\subsection{Case and Subject Description}
The study objects for RQ1 and RQ2 are based on 346 NFRs taken from 11 industrial specifications from 5 different companies for
 different application domains and of different sizes. The specifications further differ in the level of abstraction, detail, and completeness. As these specifications are confidential, we cannot give detailed information on the individual NFRs nor on the projects. However, Table~\ref{tbl:studyobjects} provides an overview of the study objects in scope of RQ1 and RQ2, their domain, and exemplary (anonymized) NFRs.  

All data classifications are performed independently by two different researchers (1st and 3rd author). Both researchers are working for more than three years in requirements engineering and model-based development research. \looseness=-1
\begin{table*}
\scriptsize
\renewcommand{\arraystretch}{1.3}
\begin{threeparttable}[b]
\caption{Overview of the study objects for RQ1 and RQ2.}
\label{tbl:studyobjects}
\centering
\begin{tabularx}{\textwidth}{llrrrX}\toprule
\bf{Spec.}&\bf{Family of Systems\tnote{1} (Domain)}&\bf{\# Reqs}&\bf{\# NFRs}&\bf{\% NFRs}&\bf{Exemplary NFR (anonymized due to confidentiality)}\\ \midrule
S1& BIS (Finance) &200&61&30.5\%&{\itshape The availability shall not be less than [x]\%. That is the current value.}\\
S2& BIS (Automotive) &177&40&22.6\%&{\itshape An online help function must be available. [It] has to be accessible in every dialog. [...]}\\
S3& BIS (Finance) &107&5&4.7\%&{\itshape  The maximal number of users that are at the same time active in the system is [x].}\\
S4& ES/BIS  (Travel Mngmt.) &38&14&36.8\% & {\itshape The [system] is used by users that are directly in contact with customers. Thus, long response times are not acceptable. The time of [x]\% of the functions within the [system]-components shall not be more than [x] seconds.}\\
S5& ES/BIS (Travel Mngmt.) &69&16&23.2\% & {\itshape It must be possible to completely restore a running configuration when the system crashes.}\\
S6& ES (Railway) &35&14&40.0\% & {\itshape The delay between passing a [message] and decoding of the first loop message shall be $\leq$ [x] seconds.}\\
S7& ES  (Railway) &122&19&15.6\% &{\itshape The collection, interpretation, accuracy and allocation of data relating to the railway network shall be undertaken to a quality level commensurate with the SIL [x] allocation to the [system] equipment.
}\\
S8& ES/BIS  (Traffic Mngmt.)&554& 128&23.1\% & {\itshape It shall be possible to install programs and configuration data separately.}\\
S9& ES (Railway)  &393& 12&3.0\% & {\itshape The [system] will have a Mean Time Between Wrong Side Failure (MTBWSF) greater than [x] h respectively a THR less than [x]/h due to the use of [a specific] platform.} \\
S10& ES (Railway) &122& 31&25.4\% & {\itshape The [system] system shall handle a maximum of [x] trains per line.}\\
S11& BIS (Facility Mngmt.) &24&6&25.0\% & {\itshape The architecture as well as the programming has to guarantee an easy and efficient maintainability.}\\
\hline
$\Sigma$ 11& & $\Sigma$ 1.841&$\Sigma$ 346&18.8\%\\
\bottomrule
\end{tabularx}
\begin{tablenotes}
            \item [1] System classes considered are BIS (Business Information systems) and ES (Embedded Systems) as well as hybrids of both.
        \end{tablenotes}
\end{threeparttable}
\end{table*}

\subsection{Data Collection \& Analysis Procedures}
We collected all requirements from the specifications that are explicitly labelled as {\itshape non-functional}, {\itshape quality}, or as one specific class of NFR, e.g. {\itshape availability}. To answer RQ1, we classified the resulting NFRs according to the following dimensions:\looseness=-1

{\bfseries Quality characteristic} from the quality model for external and internal quality (ISO\slash IEC 9126). See~\cite{iso9126} for the individual characteristics. \\
\indent {\bfseries Scope} of the NFR, i.e., \emph{system embedded in its context}, \emph{the system} itself, or a \emph{sub-system}.

For RQ2, we base our classification on one established system modeling theory~\cite{broy2012specification}. Here, we classified the NFRs according to the following fundamental dimensions (see~\cite{DBLP:journals/computer/Broy15}):

{\bfseries Modeling View}, i.e. does the NFR describe {\itshape externally visible system behavior}, {\itshape internal system behavior}, or {\itshape representational aspects}. This dimension differentiates behavior that is externally visible only, also known as black box behavior (see, e.g., NFR of S10, Table~\ref{tbl:studyobjects}), behavior that is internal to the system, also known as glass box behavior (see ex. NFR of S6, Table~\ref{tbl:studyobjects}), and representational aspects of a system (see, e.g,. NFR of S7, Table~\ref{tbl:studyobjects}).\\
{\indent \bfseries System Modeling Concept}, i.e. does the NFR describe {\itshape interface and interface behavior}, {\itshape architecture and architectural behavior}, or {\itshape state and state transition behavior}. This dimension differentiates behavior in terms of interaction over the system boundary (see, e.g., NFR of S10, Table~\ref{tbl:studyobjects}), structuring the system into a set of sub-systems with their connections and their interactions (see, e.g., NFR of S2, Table~\ref{tbl:studyobjects}), and describing the state space and state transitions of a system (see, e.g., NFR of S5, Table~\ref{tbl:studyobjects}).\\
{\indent \bfseries Modeling Theory}, i.e., with what means is the NFR described ({\itshape syntactical}, {\itshape logical}, {\itshape probabilistic}, {\itshape timed})? This dimension distinguishes between NFRs that describe syntactical structure (see, e.g., NFR of S2, Table~\ref{tbl:studyobjects}), or NFRs that describe behavior. The latter is further refined to the kind of behavior: logical, probabilistic, or timed. 

 
\section{Current Status and Preliminary Results}
At the moment of writing this paper, we completed the pre-study based on a sample and reached the validation phase of our study. So far, we analyzed 38 NFRs out of 346 (approx.\ 10\%), created the decision tree, discussed the results, and agreed on the classification. In this section, we will give a brief overview of the preliminary results and analysis.

\subsection{Preliminary Results}
\begin{figure*}
\centering
\subfigure[Modeling View]{
\includegraphics[width=0.25\textwidth]{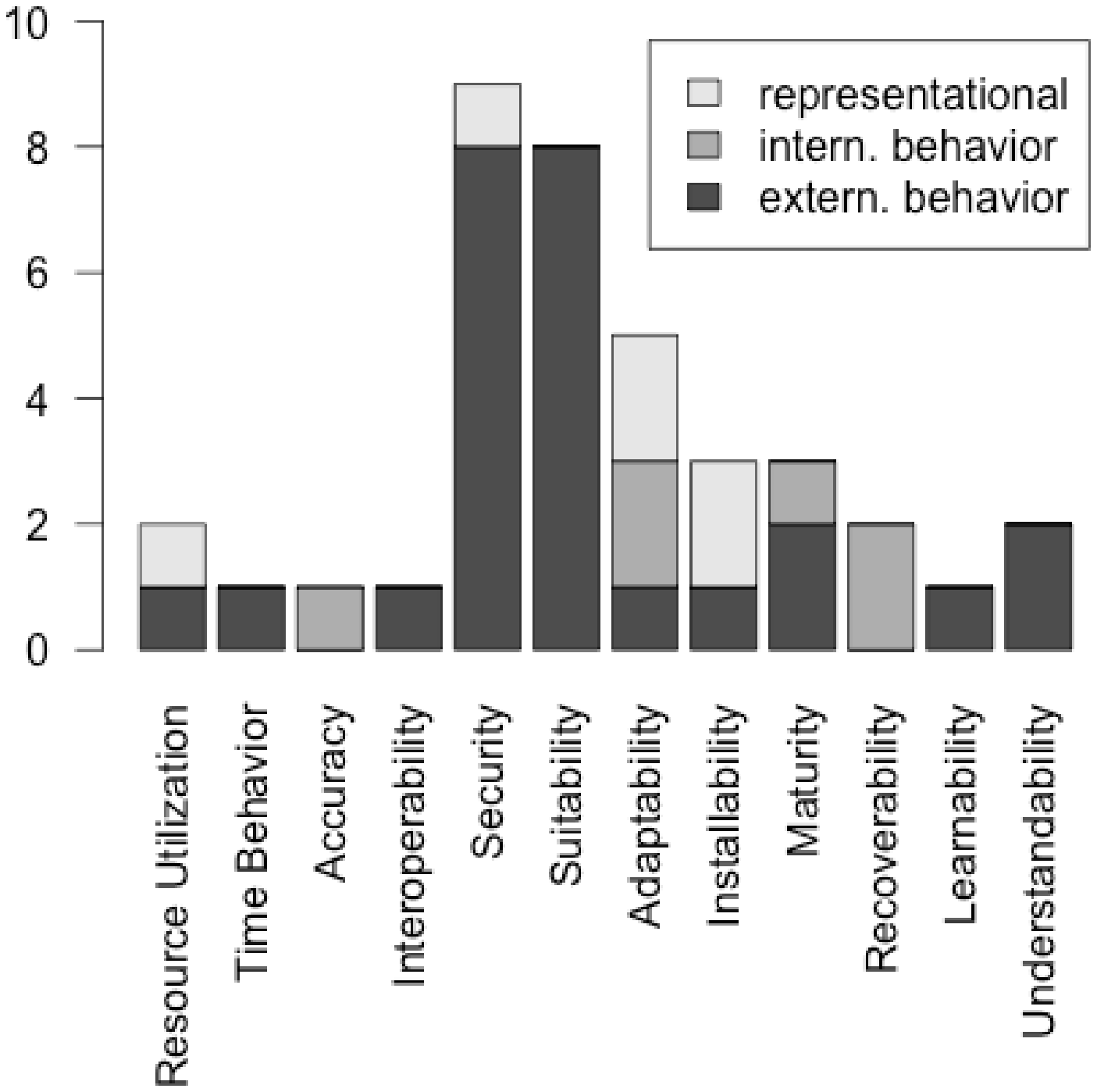}
}
\subfigure[System Modeling Concept]{
\includegraphics[width=0.25\textwidth]{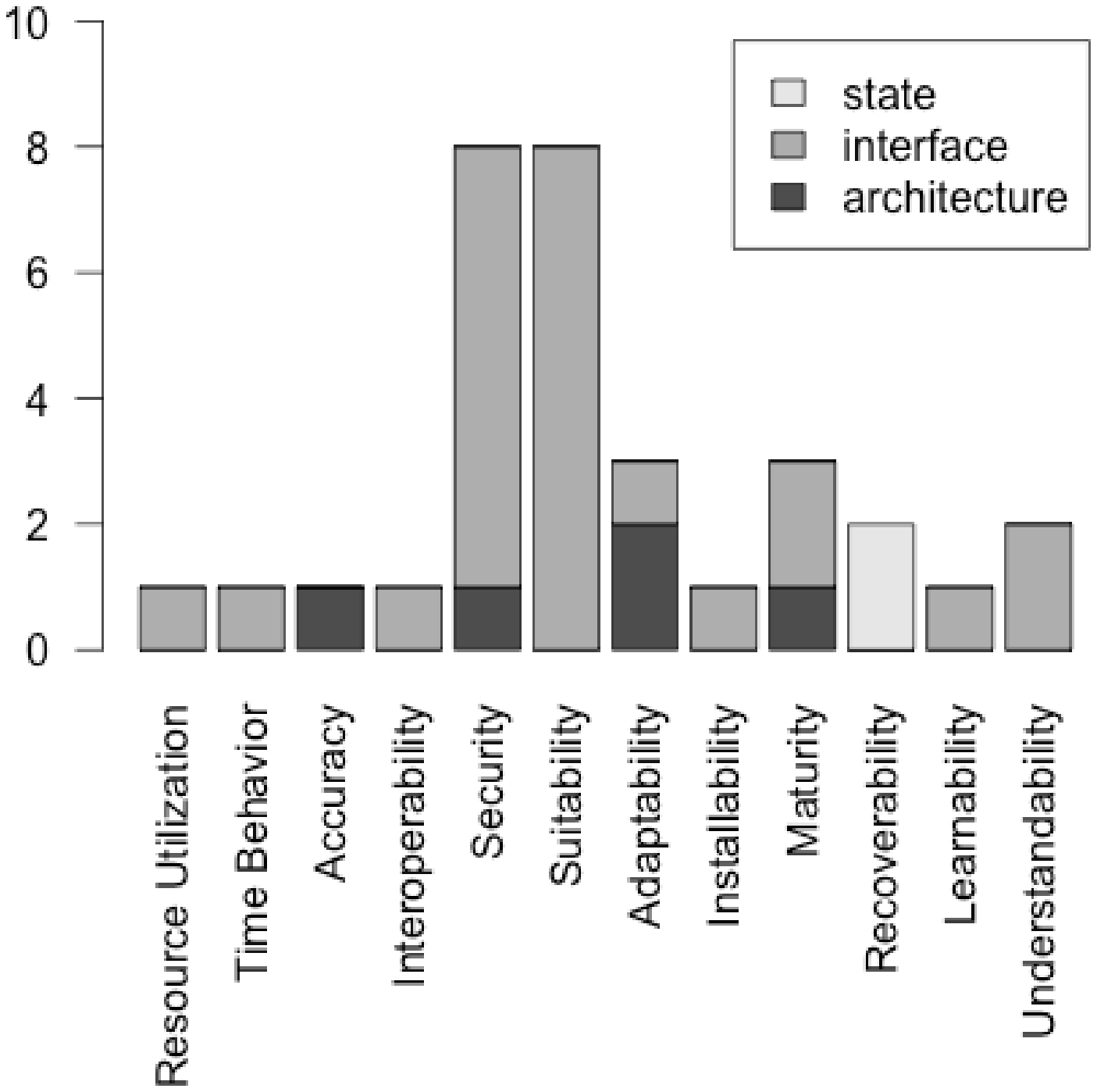}
}
\subfigure[Modeling Theory]{
\includegraphics[width=0.25\textwidth]{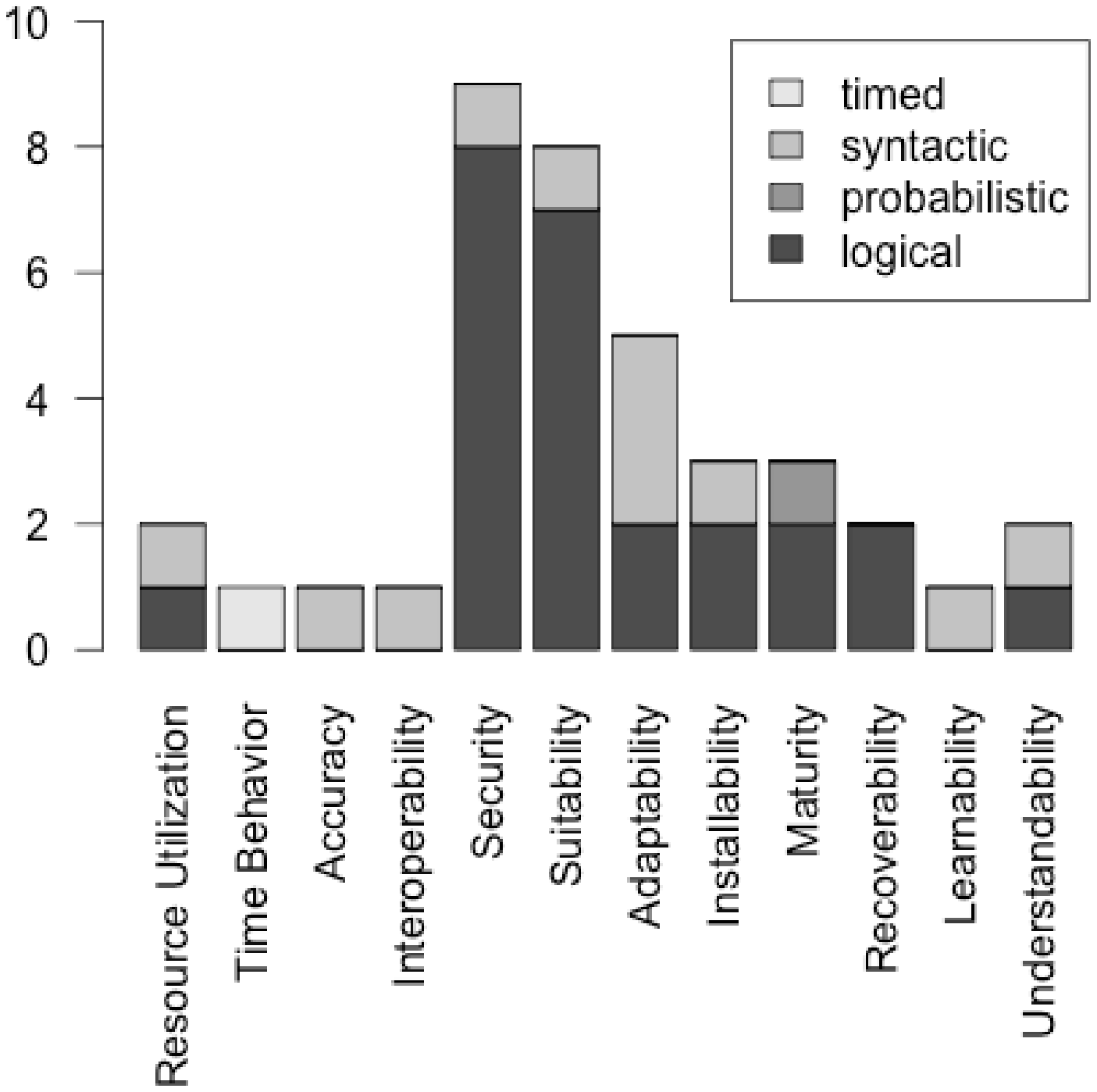}
}
\caption{Results for RQ2: Distribution of the ISO quality attributes among the system modeling dimensions.}
\label{fig:resrq2}
\end{figure*}

The results of RQ1 are shown in the following table and the results of RQ2 are shown in Figure~\ref{fig:resrq2}(a)-(c).

\begin{center}
\vspace{-0.5em}
\scriptsize
\begin{tabularx}{\columnwidth}{lRR}
\bf{Quality characteristic}& \bf{count} & \bf{percentage}\\ \midrule
Functionality - Security & 9 & 23.7\%\\       
Functionality - Suitability & 8 & 21.0\%\\
Portability - Adaptability & 5  & 13.2\%\\     
Portability - Installability  & 3 & 7.9\% \\         
Reliability - Maturity & 3 & 7.9\%\\
Reliability - Recoverability & 2& 5.3\% \\   
Usability - Understandability  & 2& 5.3\% \\
Efficiency - Resource Utilization &  2& 5.3\% \\   
Efficiency - Time Behavior & 1& 2.6\%\\  
Functionality - Accuracy & 1& 2.6\% \\
Functionality - Interoperability & 1& 2.6\% \\  
Usability - Learnability & 1& 2.6\% \\   
\end{tabularx}

\begin{tabularx}{\columnwidth}{lRR}
\bf{Scope}& \bf{count} & \bf{percentage}\\ \midrule
System in Context & 9 & 23.7\%\\       
System& 27 & 71.1\%\\
Sub-system & 2 & 5.3\%\\     
\end{tabularx}
\vspace{-2em}
\end{center}

%
%
%
%

\subsection{Preliminary Interpretation}
Concerning RQ1, one can already see that about 50\% of all NFRs in the sample are in a sub-category of {\itshape functionality}. Furthermore, within the category {\itshape functionality}, most NFRs are concerned with {\itshape security} (9)  or with {\itshape suitability} (8). The latter are classical functional requirements. Moreover within the rest, most NFRs are concerned with either portability (8) or with reliability (5). The scope of most requirements is the {\itshape system} (71.1\%), while only 23.7\% describe a functionality of the {\itshape system in its context} and only 5.3\% describe a functionality of a {\itshape sub-system}.

Concerning RQ2, we can see that in particular almost all NFRs within the {\itshape Functionality - Security} and {\itshape Functionality - Suitability} class describe externally visible behavior,  interface and interface behavior, and are described logically. Furthermore, we can see that 68.4\% of the NFRs describe externally visible behavior (15.8\% internal system behavior and 15.8\% representational aspects), 78.1\% interface and interface behavior (15.6\% architecture and architecture behavior and 6.3\% state and state transition behavior), and 78.1\% are described logically (28.9\% syntactical, 2.6\% probabilistic, and 2.6\% timed). 

In our pre-run using the sample, we did not (yet) analyze further differentiations, e.g., according to the class of systems. The results still already indicate that many NFRs seem to describe externally visible behavior, interface and interface behavior, and they are described logically. This is how classical functional requirements are specified and, furthermore, about 50\% of all NFRs in the sample are in the sub-category of {\itshape functionality}. This indicates how blurry the borders between functional requirements and NFRs actually are. 


\section{Open Issues and Threats to Validity}
\begin{table}[t]
\renewcommand{\arraystretch}{1.3}
\caption{Cohen's Kappa of 1st and 2nd prelim. classification}
\label{tbl:kappa}
\centering
\scriptsize
\begin{tabular}{lrrrrr}\toprule
\bf{Category}&{$\kappa_{v} $}(1st) & {p-val$_{v_1}$}(1st)& {$\kappa_{v_2}$}(2nd) &{p-val$_{v_2}$}(2nd)\\ \midrule
ISO\slash IEC 9126	&$0.577(10)$ 		&$\num{9.33e-05}$		&$0.505(18)$ 		&$\num{1.65e-11}$\\
Scope			&$0.0(13)$ 		&NaN				&$0.133(18)$ 		&$0.475$\\
S.M. Concept			&$0.0(11)$ 		&NaN				&$-0.0263(13)$ 	& $0.871$\\
View				&$-0.0263(13)$ 	&$0.882$				&$0.337(18)$ 		&$0.0543$\\
Theory			&$0.0(12)$ 		&NaN				&$-0.111(15)$ 	&$0.515$\\
\bottomrule
\end{tabular}
\vspace{-2em}

\end{table}

In the course of designing the study and later on during initial classifications based on the sample, we were confronted with different issues of which some still remain open. In this section, we provide an overview of those issues we consider to result in the biggest threats to the validity of our study.

{\bfseries \itshape Data Representativeness.} We see the biggest threat to the validity to be in the representativeness of the data on which we built our analysis. The concerns range from the representativeness of the way the NFRs are specified to the completeness of the data as it currently only covers the particularities of selected industrial contexts only.

{\bfseries \itshape NFR Selection.} We only collected the requirements that are explicitly labelled as non-functional or quality. With this selection procedure, some relevant NFRs may be missed or irrelevant ones may be included. To address this threat, we plan to perform the classification on the whole data set as future work (including functional and non-functional requirements). 

{\bfseries \itshape Classification Dimensions.} To answer RQ1 and RQ2, we classified our data based on multiple dimensions. One open issue concerns the validity of those dimensions themselves. The fuzziness of the dimensions manifests itself in the low inter-rater agreement and low kappa values (see Table~\ref{tbl:kappa}) which was also the reasons for us to elaborate a decision tree. Yet, another reason for the disagreement was that the NFRs were analyzed in insolation and that they often do not provide sufficient information to understand them without the necessary context. Finally, the third problem affecting the classification is given by the ISO/IEC classification itself which we took as a reference and which doesn't provide exhaustive guidance for the classification.


{\bfseries \itshape Contextualization.} The quality of our study is very much dependent on the possibility to reproduce the results, which in turn is dependent on the clearness of the context information. The latter, however, is strongly limited by NDAs that too often prevent providing full disclosure of the contexts and even the project characteristics.

\section{Conclusion}
The main goal of this short paper was to initiate discussions on the issues related on the integration of NFRs in seamless modeling in general and, in particular, discussions on open methodological issues of our study. \\
\indent To this end, we presented our overall study design which includes a multi-staged procedure to infer an empirically founded theory on specifying NFRs to support seamless modeling. Then, we discussed preliminary results from a sample (approx. 10\%) and current open issues and threats to validity of our study. The preliminary results already indicate to interesting points;  for example, many of commonly used NFR classes concern system facets in a way that shows how blurry the borders between functional and non-functional requirements are. Furthermore, we identified fields of improvement for our study, for example, the low inter-rater agreement during the classification. We conclude so far that our overall study design seems suitable to obtain the envisioned theory in the long run, but we could also show current issues that are worth discussing within the empirical software engineering community.

\bibliographystyle{IEEEtran}
\bibliography{esem_nature_short}
\end{document}